\documentclass[twocolumn, a4paper, reprint, aps, groupedaddress, superscriptaddress]{revtex4-1}
\pdfoutput=1
\usepackage{graphicx}
\usepackage{dcolumn}
\usepackage{bm}
\usepackage{blindtext}
\usepackage{hyperref}
\usepackage{placeins}
\usepackage{siunitx}
\usepackage{amsmath}
\usepackage{amsfonts}
\usepackage{amssymb}
\usepackage{mathtools}
\usepackage{physics}
\usepackage{tikz}
\usepackage[titletoc]{appendix}
\usepackage{lipsum}
\usepackage{listings}
\usepackage{xcolor}
\usepackage{float}
\usepackage{ulem}
\usepackage[export]{adjustbox}
\usepackage{resizegather}
\usepackage{caption}
\usepackage{color}
\usepackage{subfigure}
\usepackage[bottom]{footmisc}

\DeclareSIUnit \parsec {pc}
\DeclareSIUnit \h {\mbox{$h$}}

\begin{document}

\title{Thermodynamics of viscous dark energy for the late future time universe}
\author{David Tamayo}
\email{tamayo.ramirez.d.a@gmail.com}
\affiliation{Facultad de Ciencias en F\'isica y Matem\'aticas, Universidad Aut\'onoma de Chiapas, Tuxtla Guti\'errez, Chiapas, 29050, Mexico}
\affiliation{Mesoamerican Centre for Theoretical Physics, Universidad Aut\'onoma de Chiapas, Tuxtla Guti\'errez, Chiapas, 29050, Mexico}

\begin{abstract}
In this work we explore the thermodynamic aspects of dark energy for late future time universe in two different scenarios: as a perfect fluid with constant and variable equation of state parameter; and as dissipative fluid described by a barotropic equation of state with bulk viscosity in the framework of the Eckart theory and the full Israel-Stewart theory.
We explore cosmological solutions for a flat, homogeneous and isotropic universe; and we assume the late future time behavior when the dark energy dominates the cosmic evolution.
When modeled as a perfect fluid with a dynamical equation of state, $p=w(a)\rho$, the dark energy has an energy density, temperature and entropy well defined and an interesting result is that there is no entropy production even though been dynamical. 
For dissipative dark energy, in the Eckart theory two cases are studied: $\xi=const.$ and $\xi =(\beta/\sqrt{3}) \rho^{1/2}$; it is found that the entropy grows exponentially for the first case and as a power-law for the second.
In the Israel-Stewart theory we consider a $\xi =\xi_0 \rho^{1/2}$ and a relaxation time $\tau = \xi/\rho$; an analytical Big Rip solution is obtained with a power-law entropy.
In all cases a power-law relation between temperature and energy density is obtained.
In order to maintain the second law of thermodynamics theoretical constraints for the equation of state are found in the different dark energy models studied.
A barotropic dark fluid with $w<-1$ is thermodynamically difficult to support, but the overall effect of bulk viscosity in certain cases allows a phantom regime without thermodynamic anomalies.

{\bf Keywords:} Dark energy, thermodynamics, bulk viscosity.
\end{abstract}

\maketitle


\section{Introduction}

The accelerated cosmic expansion indicates the presence of a negative-pressure component to the total energy density of the universe today.   
This component is the source of the accelerated expansion and, either material fluid or geometry, is known as {\it dark energy}.
The observational evidence and theoretical consistency of dark energy is well supported; for a review see Refs. \cite{DiValentino:2021izs, Motta:2021hvl, Huterer:2017buf} and references therein.
Dark energy nowadays is a fundamental element of the standard cosmological model, however the physical mechanism behind it still a mystery.
The simplest and best-known model of dark energy is the energy of the vacuum represented as the cosmological constant added to the Einstein field equations.
The main feature of vacuum energy is that its energy density is constant in time and is spatially smooth.
The cosmological constant with cold dark matter are the key elements of the standard cosmological model or $\Lambda$CDM.

In cosmology, a perfect fluid description is adequate to model the known cosmic material components (e.g. photons, baryons, neutrinos) and also dark matter (even though we don't have a consolidated microscopic theory of dark matter, when modeled as dust, a pressureless perfect fluid, there is theoretical and observational consistency). 
Dark energy, like the other main cosmological components, as a first approach is modeled in the framework of perfect fluids in a homogeneous and isotropic expanding universe.
The perfect fluid approach applied to describe the cosmic components fit very well the cosmological observations at background and linear perturbation level; and also for some components has an underlying microscopic theory that support it.

To obtain an accelerated cosmic expansion the material content of the universe must violate the strong energy condition, $\sum_i (\rho_i + 3p_i) < 0$, where $\rho_i$ and $p_i$ are the energy density and the pressure of each component respectively. 
The total pressure must be negative.
Since baryonic and dark matter are pressureless and radiation pressure is $\rho_r/3>0$, then there must be a source of negative pressure, this source is the dark energy.
In the standard approach, the dark energy is considered as a fluid with negative pressure and barotropic equation of state, $p=w\rho$ (with $w$ constant), and value $w=-1$. 
The dark energy equation of state $p=-\rho$ implies that the energy density of dark energy is a constant, i.e., has no dynamics. 

Recent observational reconstructions of the dark energy equation of state shows that the parameter $w$ could be dynamical, i.e., depends on time (scale factor) \cite{Zhao:2017cud, Wang:2018fng} allowing to cross the phantom divide line ($w<-1$), and there are even studies that suggest a slight preference to the cosmological phantom regime $w<-1$ \cite{DAmico:2020kxu, Chudaykin:2020ghx, Yang:2021flj}, but it is not clear if it corresponds properly to a phantom fluid or to an additional physical effect which gives an effective phantom dark energy equation of state.

Dynamical dark energy, $w=w(a)$, together with the assumption of dark energy as a perfect fluid brings up some thermodynamic problems such as the positiveness of the entropy, temperature, and chemical potential implies that $w \geq -1$ which is in direct conflict with phantom dark energy \cite{Duarte:2018gmt, Silva:2013ixa}.

One way to avoid some thermodynamic problems is to suppose that the dark energy is a fluid with bulk viscosity breaking the perfect fluid hypothesis.
A fluid with bulk viscosity means that dissipative processes occur, which in cosmic fluids has the advantage of allowing the violation of the dominant energy condition ($\rho + p < 0$) without the dark energy necessarily being phantom \cite{Barrow:1988yc}.
In this case we have an effective pressure given by $p_{eff} = p +\Pi$, where $p = w\rho$ is the barotropic pressure and $\Pi < 0$ is the viscous pressure.
The possibility of explaining the accelerated expansion of the universe at the late future as an effect of the effective negative pressure due to bulk viscosity in the cosmic fluids was first considered in \cite{Zimdahl:2000zm, Balakin:2003tk}.
Viscous matter or radiation cosmologies can be mapped into the phantom dark energy scenario \cite{Velten:2013qna, Cruz:2016rqi} and also a viscous fluid is able to produce a Little Rip cosmology as a purely viscous effect \cite{Brevik:2011mm}.

A perfect fluid in equilibrium generates no entropy and no frictional heating because its dynamic is reversible and without dissipation. 
We know that real fluids behave irreversibly, and if we want to model dissipative processes we require a relativistic theory of dissipative fluids.
A classical irreversible thermodynamics was first extended from Newtonian to relativistic fluids by Eckart in 1940 \cite{Eckart:1940zz}. 
In the Eckart theory, the effective pressure of the cosmic fluid is modeled as $\Pi = -3\xi H$, where $\xi$ is a function and $H$ the Hubble parameter.
The Eckart theory has the problem that dissipative perturbations propagate at infinite speeds; this non-causal feature is its main limitation and therefore this approach could be useful to find insight from toy models but not as a realistic theory.
Nevertheless, Eckart theory has been used widely to model bulk viscosity in dark matter and dark energy models, for example, interacting viscous dark matter and dark energy \cite{Avelino:2013wea, Hernandez-Almada:2020ulm}, the possibility of crossing the phantom divide line \cite{Brevik:2005bj}, the magnitude of the viscosity to achieve this crossing using cosmological data \cite{Brevik:2015xsa}, Big Rip singularities for various forms of the equation of state parameter and the bulk viscosity \cite{Brevik:2013xxa}, and unified dark fluid cosmologies \cite{Colistete:2007xi, Avelino:2010pb, Velten:2011bg, Herrera-Zamorano:2020rdh}. 
The causal extension of the Eckart theory is the so-called Israel-Stewart theory \cite{Israel:1976tn, Israel:1979wp}.
This approach presents a better description than Eckart theory, including a casual description of dissipative processes associated to small deviations of equilibrium. 
The Israel-Stewart theory converges to the Eckart theory, when the collision time-scale in the transport equation of fluid is zero, i.e., when the bulk viscous model is noncausal and unstable.
An interesting review about viscous cosmology can be found in \cite{Brevik:2017msy} and for a dynamical analysis for a bulk viscosity dark matter model in the full Israel-Stewart formalism see \cite{Lepe:2017yvs}.
A pioneer work about dissipative processes in cosmology is \cite{Maartens:1995wt} and a highly recommended summary about relativistic fluid dynamics, dissipative relativistic fluids, applications to cosmology and astrophysics and bulk viscous perturbations is \cite{Maartens:1996vi}.
Bulk viscous dark energy has been studied in several contexts:
as phantom dark energy \cite{Cruz:2016rqi, Cruz:2017lbu, Cruz:2018arw};
dark energy with bulk viscosity observational constraints were studied in \cite{Wang:2017klo, daSilva:2018ehn, Yang:2019qza, Odintsov:2020voa};
the present acceleration of the universe as effect of bulk viscosity of self-interacting scalar field \cite{Gagnon:2011id};

An important problem in cosmology is that most of the dark energy models are able to adjust fairly well the observational data, this degeneracy hinders the tests and selection of more well-grounded dark energy models. 
The data shows only an accelerated cosmic expansion, but does not reveal the intrinsic nature of the source that causes this acceleration.
Despite unknown, dark energy should be consistent with the known laws of physics; for that reason the dark energy when modeled as an exotic fluid must satisfy the bounds imposed by the laws of thermodynamics. 
In this vein, dark energy thermodynamics has been studied by several authors: 
a remarkable theoretical treatment in the context of perfect fluids can be found in \cite{Duarte:2018gmt};
thermodynamic properties of dark energy with varying equation of state parameter $w(a)$ \cite{Saridakis:2009uu, Silva:2013ixa, Cardone:2016ewm};
and, thermodynamic properties of dark energy as a self-interacting complex scalar field \cite{Bilic:2008zz}.
Studies about the thermodynamics of dark energy with viscosity:
the validity of the generalized second law of thermodynamics in a non-flat universe in the presence of viscous dark energy \cite{Setare:2010zz}; some thermodynamic aspects of an alternative to the standard expression for bulk viscosity \cite{Disconzi:2014oda}; the thermodynamic stability analysis of non-interacting diffusive cosmic fluids with barotropic equation of state \cite{Maity:2019knj}; and, thermodynamics of viscous dark energy in the braneworld context \cite{Setare:2010zza, Nozari:2013qta}.
Following this path, in this work we investigate some general thermodynamic aspects of dark energy modeled as a perfect fluid with constant equation of state parameter $w=w_0$, dynamical dark energy $w=w(a)$, and bulk viscosity dark energy in the Eckart and Israel-Stewart theories.

This paper is organized as follows: in section \ref{general considerations} is presented the general cosmological and thermodynamic considerations used all along the paper. 
The paper is divided into two parts, the first one, dedicated to dark energy as a perfect fluid, first, section \ref{DE perfect fluid} summarizes the main results of the thermodynamics of dark energy with constant equation of state parameter, the same treatment but with dynamical dark energy is presented in \ref{DE wa}.
In second part, first in \ref{dissipative DE} are established the main ideas of a dissipative fluid in a flat, homogeneous and isotropic universe used in this work, then in section \ref{DE Eckart} is presented the dark energy thermodynamic analysis in the Eckart framework to then in section \ref{DE IS} in the Israel-Stewart.
Finally, in \ref{conclusions} we presented the conclusions. 

\section{General considerations}\label{general considerations}

In this section we present the general cosmological and thermodynamic considerations used all along the paper.
Let us consider a Friedman-Lemaître-Robertson-Walker universe (FLRW), i.e., a homogeneous, isotropic spacetime, and the flat universe case. 
We are considering the late future time universe in which the dominant component is the dark energy, $\sum_i \rho_i \approx \rho$.
In what follows we assume only one fluid as the main component of the universe, which experiments a dissipative process during cosmic evolution.
With these considerations the Friedmann equations are
\begin{eqnarray}
    H^2 &=& \frac{1}{3}\sum_i\rho_i \approx \frac{1}{3}\rho,\label{friedman 1} \\
    \dot{H} + H^2 &=& -\frac16\sum_i[\rho_i +3p_i] \approx -\frac16[\rho +3p],\label{friedman 2}
\end{eqnarray}
where the dot denotes derivatives with respect to the cosmic time, $H = \dot{a}/a$ is the Hubble parameter and we use natural units $8\pi G=c=1$.

In a FLRW cosmology the energy density conservation equation is
\begin{eqnarray}
 \dot{\rho} +3H(\rho+p)=0, \label{conservation}
\end{eqnarray}
where $\rho$ is the energy density and $p$ is the pressure. 
58

The thermodynamic assumptions are: the physical three dimensional volume of the universe at a given time can be expressed in terms of the scale factor $V=V_0 a^3(t)$ (where $V_0$ is the three dimensional volume at the present time), the internal energy of the a cosmological fluid is $U = \rho V$, and the first law of thermodynamics is expressed like
\begin{equation}\label{first law}
    T dS = dU +p dV -\mu dN,
\end{equation}
where $T$, $S$, $\mu$ and $N$ are the temperature, entropy, chemical potential and the number of particles respectively.

The temperature will be assumed as a function of the number of particle density, $n=N/V$, and the energy density, $\rho$, therefore $T=T(n,\rho)$.
The last assumption gives the following useful relation \cite{Maartens:1995wt, Maartens:1996vi}
\begin{eqnarray}\label{T(n,rho)}
    n\frac{\partial T}{\partial n} +(\rho+p)\frac{\partial T}{\partial \rho} = T \frac{\partial p}{\partial \rho}.
\end{eqnarray}
These are the main general work hypothesis of this work.

\section{dark energy as a perfect fluid with $w= constant$}\label{DE perfect fluid}

In this section we will present the main thermodynamic properties of dark energy as a barotropic perfect fluid with constant equation of state parameter.
This section is a summary of the previous work \cite{Duarte:2018gmt}.

The main considerations in this section are: a barotropic equation of state $p=w\rho$ (with the condition $w<-1/3$ to ensure the accelerated expansion) and the conservation of the number of particle current, $n^\alpha=n u^\alpha$ ($u^\alpha$ is the 4-velocity), which for a perfect fluid this is a conserved quantity $n^\alpha_{;\alpha} = 0$:
\begin{eqnarray}
    p &=& w\rho, \label{eos fp}\\
    \dot{n} &+& 3Hn = \frac{\dot{N}}{N} = 0, \label{conservation n pf}
\end{eqnarray}
where $w$ is a constant. 
The last two equations \eqref{eos fp} and \eqref{conservation n pf} in the energy density conservation equation \eqref{conservation} and the first law of thermodynamics \eqref{first law} gives:
\begin{eqnarray}
    \dot{\rho} &=& -3(1+w) H\rho, \label{conservation pf} \\
    T dS &=& Vd\rho +(1+w)\rho dV. \label{first law pf}
\end{eqnarray}
Now we can calculate some relevant thermodynamic quantities.

\subsection{Temperature and energy density}

Assuming that the energy density is a function of the temperature and volume, $\rho = \rho(T, V)$, then
\begin{eqnarray}
d\rho &=& \frac{\partial \rho}{\partial T}dT + \frac{\partial \rho}{\partial n}dn, \\
\Rightarrow\; \frac{d\rho}{da} &=& \frac{\partial \rho}{\partial T}\frac{dT}{da} + \frac{3n}{a}\frac{\partial \rho}{\partial n},
\end{eqnarray}
combining the last equation with the perfect fluid energy density conservation equation \eqref{conservation pf} we have
\begin{eqnarray}
    (1+w)\rho = -\frac{a}{3}\frac{\partial \rho}{\partial T}\frac{dT}{da} -n\frac{\partial \rho}{\partial n}, 
\end{eqnarray}
using now the relation of the temperature \eqref{T(n,rho)} and the equation of state \eqref{eos fp}, we obtain a relation for the temperature
\begin{eqnarray}\label{dlnT}
    \frac{dT}{T} = -3w \frac{da}{a} = \frac{d\rho}{\rho} +3 \frac{da}{a}.
\end{eqnarray}
Integrating the last equation we have 
\begin{eqnarray}\label{Temp rho U}
    \frac{T}{T_{0}} = \frac{\rho}{\rho_{0}}a^3 = \frac{U}{U_{0}}.
\end{eqnarray}
The last expression relates temperature, energy density and internal energy in the expected way for a perfect fluid, the temperature is directly proportional to the internal energy.

To calculate the energy density in terms of the scale factor we must integrate directly the perfect fluid energy density conservation equation \eqref{conservation pf} to obtain the well-known expression
\begin{eqnarray}\label{rho a pf}
\rho = \rho_{0}\,a^{-3(1+w)}.
\end{eqnarray}
This is the usual scale factor power-law energy density.
Let us remind that in the $\Lambda$CDM case, for $w=-1$, which gives $\rho = \rho_{0}$, the energy density of the dark energy is a constant.
When, $-1< w<-1/3$, we have a dark energy that dilutes as the scale factor grows (but slower than dust or radiation).
And, for the phantom case, $w<-1$, we have $\rho = \rho_{0}\,a^{r}$ where $r$ is a positive constant, in this case the energy density grows as the scale factor increase, and consequently the temperature and internal energy too.

To calculate the energy density in terms of the temperature we use \eqref{dlnT} and \eqref{conservation pf}
\begin{eqnarray}\label{rho temp pf}
    \frac{dT}{T} &=& -3w \frac{da}{a} = \frac{w}{1+w} \frac{d\rho}{\rho} \label{dT/T},\\
    \Rightarrow \rho &=& \Tilde{\rho}\,T^{\frac{1+w}{w}}, \label{rho T}
\end{eqnarray}
where $\Tilde{\rho}$ is a constant dependent of $w$. 
We obtain a power-law relation between temperature and energy density.

\subsection{Entropy}

Using the Gibbs relation \eqref{first law pf} and the perfect fluid energy density conservation equation \eqref{conservation pf} we can easily calculate a relation for the entropy
\begin{eqnarray}
T dS &=& (1+w)\rho dV +vd\rho \nonumber\\
&=& U[d\ln \rho +3(1+w)d\ln a] = 0 \nonumber\\
&\Rightarrow& S=constant.
\end{eqnarray}
The energy density conservation equation and first law of the thermodynamics for a perfect fluid with $w$ constant in FLRW imply that the entropy is a constant, i.e. is adiabatic. 
There is no entropy produced within the system (no friction, viscous dissipation, etc.) by the work done due to the cosmic expansion. 
Notice that this result is independent of the value of $w$.

Lets calculate the entropy, in the perfect fluid case the entropy of a cosmic fluid can be calculated from the well-known Euler relation:
\begin{equation}\label{euler}
    U = T S -p V +\mu N,
\end{equation}
we can easily calculate the entropy taking into account the relation $U/U_0 = T/T_0$ from \eqref{Temp rho U}, with this we have
\begin{equation}\label{entropy w}
    S = (1+w)\frac{U}{T} -\frac{\mu N}{T}= (1+w)\frac{\rho_{0} V_0}{T_{0}} -\frac{\mu N}{T}.
\end{equation}
For a perfect fluid the entropy $S$ and the number of particles $N$ are constants, consequently the ratio $\mu/T$ must be constant.
We can define $\mu/T = \mu_0/T_0$ and by the second law of thermodynamics, $S\geq0$, then we have the following relation
\begin{eqnarray}\label{w restriction mu}
    w \geq -1 +\frac{\mu_0 n_0}{\rho_0}.
\end{eqnarray}
We have three options:
First $\mu_0=0$, in this case we have $w\geq-1$ which is the phantom divide line restriction, that is, the phantom-like behavior of the dark energy fluid is forbidden. 
For $\mu_0 > 0$, then $w$ has a minimal value that does not reach the cosmological constant point $w > -1$ and again a phantom regime is forbidden. 
Finally, if the chemical potential is negative $\mu_0 < 1$ then values of $w<-1$ are allowed, i.e, phantom dark energy. 
In \cite{Silva:2013ixa} the authors examine these cases for several dark energy models.

\section{dark energy as a perfect fluid with $w= w(a)$}\label{DE wa}

In this section we will study dynamical dark energy as a perfect fluid with the same background hypothesis of the previous section; the conservation of particle equation \eqref{conservation n pf}, the perfect fluid energy density conservation equation \eqref{conservation pf} and the first law of thermodynamics \eqref{first law pf}, hold in this case. 
But with the difference that now the equation of state has the form
\begin{eqnarray}\label{eos wa}
    p = w(a)\rho.
\end{eqnarray}
The core idea of dynamical dark energy is that the equation of state parameter is a variable quantity, $w(a)$, since it is constructed from two variable quantities.

\subsection{Temperature and energy density}

For a dynamical dark energy the relation $U/U_0 = T/T_0 = (\rho/\rho_0) a^3$ \eqref{Temp rho U} still holds.
When we consider $p=w(a)\rho$ the equation \eqref{dT/T} cannot be integrated immediately, it is necessary to know the explicit dependence with the scale factor of the equation of state parameter.
\begin{eqnarray}
    \frac{dT}{T} &=& -3w(a)\frac{da}{a},\\
    \Rightarrow\; T &=& T_0 \exp \left[-3\int w(a)\frac{da}{a}\right] = T_0 \mathcal{E}. \label{temp wa}
\end{eqnarray}
To calculate the temperature in terms of the energy density we may solve the equation of $T(n,\rho)$ \eqref{T(n,rho)} in this particular case
\begin{eqnarray}
    n\frac{\partial T}{\partial n} +[1+w(a)]\rho\frac{\partial T}{\partial \rho} = T w(a)
\end{eqnarray}
whose solution using the method of characteristics yields
\begin{eqnarray}\label{temp rho F}
    T = \rho^{\frac{w}{w+1}}\, F(\rho^{\frac{1}{w+1}}/n),
\end{eqnarray}
in which $F$ is an arbitrary function.
Notice that, for a the particular case of $w=const.$, from equation \eqref{rho a pf} we have $\rho a^{3(1+w)} = \rho_0 =const.$ and from \eqref{conservation n pf} we have $na^3=n_0=const.$; then, $\rho^{\frac{1}{w+1}}/n =const.$ which imply that $ F(\rho^{\frac{1}{w+1}}/n) =const.$ recovering the previous result $\rho=\rho_0 T^{(1+w)/w}$ of the equation \eqref{rho T}.

To calculate the energy density in terms of the scale factor we integrate the energy density conservation equation \eqref{conservation}.
\begin{eqnarray}
\rho = \rho_{0}\,a^{-3} \exp \left[-3\int w(a)\frac{da}{a}\right] = \rho_{0}\,a^{-3} \mathcal{E}. \label{rho wa}
\end{eqnarray}
Again, it is necessary to know the explicit dependence with the scale factor of the equation of state parameter in order to obtain a closed $\rho(a)$.

\subsection{Entropy}

Using the first law of thermodynamics equation \eqref{first law}, the energy density conservation equation \eqref{conservation} and the equation of state \eqref{eos wa} we can easily calculate
\begin{eqnarray}
T dS &=& U[d\ln \rho +3(1+w(a))d\ln a] = 0\\
&\Rightarrow& S=S_0=constant.
\end{eqnarray}
There is no entropy production in a universe filled with perfect fluid dark energy despite having a dynamical equation of state, $w(a)$. 
This is an interesting result. 

If we use the Euler relation $U = T S -p V +\mu N$, the temperature relation \eqref{temp wa}, the equation of state \eqref{rho wa} and $N=nV=n_0 V_0$, then we have
\begin{eqnarray}
    S_0 &=& [1+w(a)]\frac{\rho_{0} V_0}{T_{0}} - \frac{\mu n_0 V_0}{T_0 \mathcal{E}},\\
    \Rightarrow\; w(a) &=& -1 + \frac{n_0}{\rho_0}\left(\frac{S_0 T_0}{V_0 n_{0}} + \frac{\mu}{\mathcal{E}} \right).
\end{eqnarray}
If the chemical potential is null, $\mu=0$, or if it is of the form $\mu = \mu_0\, \mathcal{E}$, then there is no dynamical dark energy, $w(a)=w= const.\geq -1$.
A positive chemical potential, $\mu>0$, results in a $w(a)>-1$ avoiding the cosmological constant case.
Finally, for $\mu<0$ and $|\mu| < \frac{S_0 T}{V_0 n_0}$, again we have $w(a)>-1$; and for $|\mu| > \frac{S_0 T}{V_0 n_0}$, we have the phantom case $w(a)<-1$ \cite{Silva:2013ixa}.
This model allows the phantom dark energy with positive definite temperature and entropy in certain cases of negative chemical potential.

\section{Dissipative dark energy}\label{dissipative DE}

A common feature of many of dark energy models is the assumption that it can be modeled as a perfect fluid.
As stated before, a perfect fluid generates no entropy and no frictional heating because its dynamics is reversible and without dissipation. 
This approach works quite well in standard cosmology, but the observational reconstructions of the dark energy equation of state suggest the possibility of phantom dark energy which is incompatible with the perfect fluid hypothesis at thermodynamic level. 
Real fluids behave irreversibly, for this reason, if we want to maintain the fluid hypothesis to dark energy and ensure its thermodynamic compatibility, it is interesting to consider dissipative processes like bulk viscosity. 
Qualitatively, the bulk viscosity can be interpreted as a macroscopic consequence coming from the frictional effects within the fluid. 

For a dissipative fluid, the particle 4–current will be taken to be of the same form as $n^\alpha_{;\alpha}=0$, this corresponds to choosing an average 4–velocity in which there is no particle flux, known as the particle frame. 
At any event in spacetime, it is considered that the thermodynamic state of the fluid is close to a fictitious equilibrium state at that event. 
The local equilibrium scalars are denoted with a bar: $\bar{n}$, $\bar{\rho}$, $\bar{p}$, $\bar{S}$, $\bar{T}$ and the local equilibrium 4–velocity is $\bar{u}^\mu$.
In the particle frame, it is possible to choose $\bar{u}^\mu$ such that the number and energy densities coincide with the local equilibrium values, while the pressure in general deviates from the local equilibrium pressure:
\begin{eqnarray}\label{thermo scalars}
    n=\bar{n}, \quad \rho=\bar{\rho}, \quad p=\bar{p}+\Pi,
\end{eqnarray}
where $\Pi$ is the bulk viscous pressure. 
We are considering in the last equations \eqref{thermo scalars} that the thermodynamical system is in the near equilibrium regime, if the fluid is out of equilibrium as a result of dissipative effects then there is no unique average 4–velocity spoiling the formalism \cite{Maartens:1995wt}.
In the presented case, the near equilibrium condition occurs when the local equilibrium pressure $p$ is dominant over the viscous pressure (which encodes deviations from equilibrium), i. e., when the condition 
\begin{equation}\label{near eq cond}
\left|\frac{\Pi}{p}\right| \ll 1.    
\end{equation}
is satisfied. 
This condition is the same for both non causal and causal approaches.
For notation, from now we will drop the bar of the thermodynamic variables and write the pressure as $p_{eff} = p +\Pi$, where $p$ is the usual barotropic pressure $p =w\rho$.
The energy density conservation equation in this case is 
\begin{equation}\label{conservation dissipative}
    \dot{\rho} + 3H(\rho +p) + 3H\Pi =0.
\end{equation}
From the first law of thermodynamics \eqref{first law} and remembering that for the energy $U=\rho V$, volume $V=V_0 a^3$ and number of particle density $n=N/V$, can be derived an expression for the entropy in terms of the bulk viscous pressure
\begin{eqnarray}\label{entropy viscous}
    nT \frac{dS}{dt} = -3H \Pi. 
\end{eqnarray}
The above equation sets a strong general constraint in this framework: due to $n$ and $T$ are positive, and $H$ in an expanding universe is positive too to ensure a non-negative entropy production (second law of thermodynamics) the bulk viscous pressure necessarily must be $\Pi \leq 0$.

\section{Eckart bulk viscosity dark energy model}\label{DE Eckart}

The simplest approach to treat bulk viscosity in cosmology is the Eckart theory \cite{Eckart:1940zz}.
This approach has some important limitations, the main one is that it is a noncausal approach to dissipative processes. 

In this framework the bulk viscosity pressure is given by 
\begin{eqnarray}\label{Eckart}
    \Pi = -3\xi H,
\end{eqnarray}
where the bulk viscosity depends on the function $\xi$ and the Hubble parameter. 
This formalism has been widely used at background level. 
Parameter constraints from observational data of the Eckart bulk viscosity dark energy can be found in Refs. \cite{Wang:2017klo, daSilva:2018ehn}; in both works the main results are that the Bayesian evidence show that the Eckart viscous dark parameters are small and statistically indistinguishable from $\Lambda$CDM with the current data and the effective EoS is slightly phantom.
First, in \cite{Wang:2017klo} the authors study three Eckart viscous dark energy models: $\tilde {p}_{de} = -\rho_{de} - 3\eta H^2$ and $\tilde {p}_{de} = \omega \rho_{de} - 3\eta H^2$, both combined with presureless matter and curvature. Note that $\xi = \eta H$ and when it is only considered dark energy, it is equal to model 2 of this work. Using  combined CMB +SNIa +BAO +Cosmic chronometer +gravitational lensing data they found the 2$\sigma$ upper bounds of the parameter $\eta < 0.003$, also for the $\omega$DE model they get $\omega = -1.001^{+0.012}_{-0.011}$.
In \cite{daSilva:2018ehn} the authors make a Bayesian analysis using CMB +SNIa +BAO +Cosmic chronometer data for three models: $\tilde {p}_{de} = \omega\rho_{de} -3\xi_0 H^2$ model I (when it is only considered dark energy, it is equal to model 2 of this work), $\tilde {p}_{de} = -\rho_{de} -3\xi_0 \sqrt{\rho_{de}}$ model II, and $\tilde {p}_{de} = \omega \rho_{de} -3\xi_0 H$ model III (when it is only considered dark energy, it is equal to model 1 of this work). 
They define the parameter $\tilde{\xi} = (8\pi G/H_0)\xi_0$ and the general parameter constraints are: $\tilde{\xi} = -0.0097 \pm 0.013$ for model I, $\tilde{\xi} = -0.012 \pm 0.019$ for model II and $\tilde{\xi} = -0.002 \pm 0.008$ for model III.
Draws attention the negative sign of $\tilde{\xi}$, this implies $\Pi>0$ which is forbidden, however the error statistically allows the results.
Finally, they found $\omega \sim -1.05$, slightly phantom too.

From \eqref{conservation dissipative}, the energy density conservation equation is 
\begin{equation}\label{bv conservation}
    \dot{\rho} + 3H(1+w)\rho -9H^2 \xi =0.
\end{equation}
First, we can sketch some general features without determining the functional form of $\xi$.
From the Friedmann equations \eqref{friedman 1} and \eqref{friedman 2} we have
\begin{eqnarray}\label{friedmann 2 Eckart}
    2\dot{H} +3(1+w)H^2 = 3\xi H.
\end{eqnarray}
Note that for the case without viscosity $\xi=0$ and $w=-1$ the last equation boils down into $\dot{H}=0$ which is the de Sitter case.
Now, assuming $w=const.$, integrating the last equation calculate $H$ as a function of the bulk viscosity we have \cite{Cataldo:2005qh}. 
\begin{eqnarray}
    H(t) = \frac{\exp\left[\frac32\int\xi(t)dt\right]}{C +\frac32(1+w)\int\exp\left[\frac32\int\xi(t)dt\right]dt},
\end{eqnarray}
where $C$ is an integration constant. 
Integrating again, for $w \neq -1$, we obtain an expression for the scale factor:
\begin{eqnarray}\label{Eckart scale factor}
    a(t) = D\left(C +\frac32(1+w)\int\exp\left[\frac32\int\xi(t)dt\right] dt\right)^{\frac{2}{3(1+w)}}
\end{eqnarray}
where $D$ is a another integration constant. 

\subsection{Temperature and energy density}

As in the perfect fluid case, to calculate the temperature in terms of the energy density we have to solve the equation \eqref{T(n,rho)} whose solution is \eqref{temp rho F}, $T = \rho^{\frac{w}{w+1}}\, F(\rho^{\frac{1}{w+1}}/n)$, in which $F$ is an arbitrary function.
To find an approximation lets define $z=\rho^{\frac{1}{w+1}}/n$ and differentiate 
\begin{eqnarray}\label{temp eckart}
    \dot{T} = \left(\frac{w}{w+1}\right) T \frac{\dot{\rho}}{\rho} + \rho^{\frac{w}{w+1}} \frac{dF}{dz} \dot{z},
\end{eqnarray}
using the equations \eqref{conservation} and \eqref{conservation n pf} in $\dot{z}$, we have
\begin{eqnarray}
    \dot{z} = \frac{\rho^{\frac{w}{w+1}}}{n}\left[\left(\frac{1}{w+1}\right)\frac{\dot{\rho}}{\rho} -\frac{\dot{n}}{n}\right] =0.
\end{eqnarray}
Then the equation \eqref{temp eckart} reduces to \eqref{rho temp pf} whose solution has already been calculated, $\rho = \Tilde{\rho}\,T^{\frac{1+w}{w}}$, which is the same relation of energy density and temperature of the perfect fluid case.

In order to calculate specific expressions for the energy density and temperature lets analyze two particular popular proposals for $\xi$.\\ 

{\bf Model 1 $\xi(t) = \xi_0 =const.$:} this is simplest Eckart bulk viscosity model.
Solving \eqref{friedmann 2 Eckart} for model 1 we have $2\dot{H} +3(1+w)H^2 = 3\xi_0 H$, with the condition that at the present time $t_0$, the Hubble parameter $H(t_0) =H_0$.
\begin{eqnarray}\label{H Eckart 1}
    H(t) = \frac{H_0 \xi_0\, e^{\frac32 \xi_0 (t-t_0)}}{H_0(1+w) \, [e^{\frac32 \xi_0 (t-t_0)}-1] +\xi_0}
\end{eqnarray}
Integrating again with the condition $a(t_0)=1$ we obtain the scale factor
\begin{eqnarray}\label{a Eckart 1}
    a(t) = \left[1 +\frac{H_0(1+w)}{\xi_0}(e^{\frac32 \xi_0 (t-t_0)}-1)\right]^{\frac{2}{3(1+w)}},
\end{eqnarray}
from which can be calculated the energy density in terms of the scale factor
\begin{eqnarray}\label{rho Eckart 1}
    \rho(a) = \rho_0 \left[W +(1-W) a^{-\frac{3(1+w)}{2}}\right]^2,
\end{eqnarray}
with the constant $W = \xi_0/(H_0(1+w))$. 
The near equilibrium condition \eqref{near eq cond} for $\xi(t) = \xi_0$ implies $|\xi_0|\ll H_0$, and consequently for $-1<w<-1/3$ imply that $W$ is small.

Notice that the energy density has the form $\rho = c_1 + c_2 a^{-3(1+w)} +c_2 a^{-3(1+w)/2}$, where $c_i$ are constants.
The first term behaves as the cosmological constant, the second as a perfect fluid in $\Lambda$CDM and the last as a slow diluting fluid corresponding to the bulk viscosity effect.\\

{\bf Model 2 $\xi=(\beta/\sqrt{3}) \rho^{1/2}$:} in \cite{Barrow:1986yf} was first assumed that the viscosity has a power-law dependence upon the energy density, $\xi=\alpha \rho^{s}$, where $\alpha>0$ and $s$ are constant parameters.
In model 2 is a great difficulty to obtain easily manageable solutions to the main equations, only some particular results have been found. 
In the special case where the bulk viscosity coefficient takes the form $\xi(\rho) \propto \rho^{1/2}$, a Big Rip singularity solution was obtained in this formalism for late future times FLRW flat universe filled with only one barotropic fluid with bulk viscosity \cite{Cataldo:2005qh}. 
For $s = 1/2$ and $\alpha=\beta/\sqrt{3}$ in $\xi=\alpha \rho^{s}$, the equation \eqref{friedmann 2 Eckart} simplifies into
\begin{eqnarray}
    2\dot{H}+3H^2[1+w-\beta] = 0,
\end{eqnarray}
which can be easily integrated to calculate the Hubble function and the scale factor
\begin{eqnarray}
    H(t) &=& H_0 \left[1 +\frac{3H_0}{2}(1+w-\beta)(t-t_0)\right]^{-1},\label{H Eckart 2} \\
    a(t) &=& \left[1 +\frac{3H_0}{2}(1+w-\beta)(t-t_0)\right]^{\frac{2}{3(1+w-\beta)}}, \label{a Eckart 2} 
\end{eqnarray}
With this, we can calculate the energy density
\begin{eqnarray}
    \rho(a) &=& \rho_0\, a^{-3(1+w-\beta)}.
\end{eqnarray}
The energy density is a power-law of the scale factor.
It is important to notice that for $w=-1$ dark energy the exponent is positive and consequently is an increasing energy density and temperature. 
The near equilibrium condition \eqref{near eq cond} in this model $\xi(t) = (\beta/\sqrt{3})\rho^{1/2}$ implies $\beta \ll |w|$.
For values $w\neq -1$, some combinations of the parameters can result into different evolutions of the energy density.
For a decreasing energy density (diluted by the cosmic expansion like occurs with ordinary matter and radiation) it is needed negative powers, i.e., $\beta < w+1$ must hold; observations show that $w+1$ is close to zero (and positive or negative), so for a decreasing energy density the conditions $w+1>0$ and $\beta \ll 1 $ are necessary. 
Phantom dark energy, $w+1<0$, or relatively big values of $\beta$, always yields $\beta > w+1$, the increasing energy density.

\subsection{Entropy}

We calculate the entropy using the equation \eqref{entropy viscous}, substituting the Eckart the bulk viscosity pressure, $\Pi = -3\xi H$, we have
\begin{eqnarray}\label{dS Eckart}
    \frac{dS}{dt} = \frac{9H^2}{nT} \xi(t). 
\end{eqnarray}
As in the previous subsection we will treat two cases, first we calculate the thermodynamic variables $n$ and $T$, substitute in \eqref{dS Eckart} and then integrate.\\

{\bf Model 1 $\xi(t) = \xi_0 =const.$:} first note that a growing entropy in time, $dS/dt>0$, immediately imply that the bulk viscosity parameter is a positive constant $\xi_0 >0$.
Considering only this option the Hubble function is given by equation \eqref{H Eckart 1}.

The particle density $n = n_0 a^{-3}$ can be easily calculated using \eqref{a Eckart 1}
\begin{eqnarray}
    n(t) = n_0 \left[1 +\frac{H_0(1+w)}{\xi_0}(e^{\frac32 \xi_0 (t-t_0)}-1)\right]^{-\frac{2}{1+w}}.
\end{eqnarray}
And the temperature $T = \tilde{T}\, \rho^{\frac{w}{1+w}} $
\begin{eqnarray}
   T(t) = \tilde{T}\,\rho_0^{\frac{w}{1+w}} \left[\frac{\xi_0\, e^{\frac32 \xi_0 (t-t_0)}}{H_0(1+w) \, [e^{\frac32 \xi_0 (t-t_0)}-1] +\xi_0}\right]^{\frac{2w}{1+w}}.
\end{eqnarray}
Substituting these two previous expressions in \eqref{dS Eckart} we have a differential equation for the entropy that depends only on time
\begin{eqnarray}\label{dS Eckart 1}
    \frac{dS}{dt} = \frac{3 \xi_0 \rho_0^{\frac{1}{1+w}}}{n_0 \tilde{T}} \, e^{\frac{3\xi_0}{1+w}(t-t_0)}.
\end{eqnarray}
A quick examination of the last equation gives a constriction to the equation of state parameter $w$, in order to ensure the second law of thermodynamics, it must satisfy $w>-1$, i.e. the phantom case is forbidden for a barotropic fluid.
Another possibility is the to allow unusual assumptions like a negative temperature $\tilde{T}<0$ in order to switch the sign and yield an increasing rate of entropy.
The option of negative temperature in the phantom regime was examined in \cite{Saridakis:2009uu}.
 
The equation \eqref{dS Eckart 1} can be integrated easily if we assume $w= constant$,
\begin{eqnarray}\label{S Eckart 1}
    S(t) = S_0 +\frac{(1+w) \rho_0^{\frac{1}{1+w}}}{n_0 \tilde{T}}\, [e^{\frac{3\xi_0}{1+w}(t-t_0)}-1],
\end{eqnarray}
where $S_0$ is the entropy at the present time.
The entropy in this particular model \eqref{S Eckart 1} grows exponentially.\\

{\bf Model 2 $\xi=(\beta/\sqrt{3}) \rho^{1/2}$:} Repeating the same procedure of the previous case, calculating the particle density, temperature and substituting in \eqref{dS Eckart}; first we obtain the differential equation for the entropy
\begin{eqnarray}\label{dS Eckart 2}
    \frac{dS}{dt} &=& \frac{\beta\rho_0^{\frac{2+w}{1+w}}}{H_0 n_0 \tilde{T}} \left[1 +\frac{3H_0}{2}(1+w-\beta)(t-t_0)\right]^{-1 +2\delta} \\
    \delta &=& -1 +\frac{w}{1+w} +\frac{1}{1+w-\beta}.
\end{eqnarray}
Let's quickly examine equation \eqref{dS Eckart 2} before solving it.
First, clearly for $\beta=0$ (no bulk viscosity) then $dS/dt=0$, i.e. the entropy is a constant in time as expected for a perfect fluid.
For small values of the bulk viscosity function parameter, $0<\beta \ll 1$, we have $\delta \approx 0$, which implies that $dS/dt \propto [1 +\frac{3H_0}{2}(1+w-\beta)(t-t_0)]^{-1}$, the condition of increasing entropy $dS/dt>0$ requires that $w\geq -1$.

Integrating we obtain the expression of the entropy
\begin{eqnarray}
    S(t) &=& S_0 + \frac{(1+w)\rho_0^{\frac{1}{1+w}}}{n_0 \tilde{T}} \nonumber\\
    &\times& \left[\left(1 +\frac{3H_0}{2}(1+w-\beta)(t-t_0)\right)^{2\delta}-1\right],
\end{eqnarray}
again, $S_0$ is the entropy at the present time.
The entropy in model 2 is a power-law.  
An increasing entropy condition imposes $\delta>0$, from which we have two cases: for $w+1>0$ (quintessence) it is required $\beta<1+w$ and as we know the equation of state is close to the cosmological constant case, i.e. $0< w+1 \ll 1$ then $\beta$ has to be very small; the second case is $w+1<0$ (phantom) which implies $1+w<\beta$ which it is easily fulfilled. 

\section{Israel-Stewart bulk viscosity dark energy model}\label{DE IS}

The Israel-Stewart theory provides a better description than the Eckart theory.
It is a causal and stable theory of thermal phenomena in the presence of gravitational fields.
This theory besides solving the non-causal problem of the Eckart theory, enrich the framework including new features like the entropy has terms of second order in the dissipative variables and incorporates transient phenomena on the scale of the mean free path/time, outside the quasi–stationary
regime of the classical theory.
In the Israel–Stewart theory we have the same Friedmann equations with an equation for the causal evolution of the bulk viscous pressure given by
\begin{eqnarray}\label{transport IS}
\tau \dot{\Pi} + \Pi = -3\xi H -\frac12 \tau \Pi \left(3H +\frac{\dot{\tau}}{\tau} -\frac{\dot{\xi}}{\xi} -\frac{\dot{T}}{T}\right)
\end{eqnarray}
The last equation is known as the transport equation of the viscous pressure $\Pi$.
Where $\tau$ is the relaxation time for bulk viscous effects (in the limit $\tau = 0$ the theory is noncausal), $\xi$ the bulk viscosity coefficient and $T$ the temperature.

It is clear that the Israel-Stewart theory is much more complex than the Eckart theory.
In addition to having to propose or infer a function for $\xi$ and for the relaxation time $\tau$, we must know the temperature $T$ and solve a differential equation involving these physical quantities.
Let's see how we can make a proposal and look under which conditions an analytical solution of equation \eqref{transport IS} can be obtained.

First for the temperature, following \cite{Cruz:2016rqi, Maartens:1995wt, Maartens:1996vi}, we assume that the temperature depends only on the particle number an energy density, $T=T(n,\rho)$, with this 
\begin{eqnarray}
    dT = \left(\frac{\partial T}{\partial \rho}\right)_n d\rho + \left(\frac{\partial T}{\partial n}\right)_\rho dn, \\
    \Rightarrow \; \frac{\dot{T}}{T} = \frac{1}{T}\left(\frac{\partial T}{\partial \rho}\right)_n \dot{\rho} + \frac{1}{T}\left(\frac{\partial T}{\partial n}\right)_\rho \dot{n}. \label{dotT/T 1}
\end{eqnarray}
The conservation equations of energy density and particle density in the Israel-Stewart theory are
\begin{eqnarray}\label{conservation eq IS}
\dot{\rho} &=& -3H(\rho+p +\Pi),\\
\dot{n} &=& -3Hn
\end{eqnarray}
substituting both expressions in \eqref{dotT/T 1} and using \eqref{T(n,rho)} we have
\begin{eqnarray}\label{dotT/T 2}
    \frac{\dot{T}}{T} = -3H\left[\left(\frac{\partial p}{\partial \rho}\right)_n +\frac{\Pi}{T} \left(\frac{\partial T}{\partial \rho}\right)_n\right].
\end{eqnarray}
If $w$ is not a function of $\rho$, then $(\frac{\partial p}{\partial \rho})_n =w$ (which $w$ is not necessarily a constant). 
Taking this into account this, it is easy to prove that the expression for the temperature 
\begin{eqnarray}\label{Temp IS}
T &=& \tilde{T} \rho^{\frac{w}{1+w}}, 
\end{eqnarray}
is solution of \eqref{dotT/T 2}.
Which is the same as for the perfect fluid case and Eckart bulk viscosity. 

For $\xi$, we assume a power-law for the bulk viscosity in terms of the energy density of the fluid \cite{Barrow:1986yf},
\begin{eqnarray}\label{xi}
\xi &=& \xi_0 \rho^s, 
\end{eqnarray}
where $s$ is a constant arbitrary parameter and $\xi_0$ a positive constant. 

And finally we follow \cite{Cataldo:2005qh} in which a simple relation between $\tau$ and $\xi$ is proposed 
\begin{eqnarray}\label{tau}
\tau &=& \frac{\xi}{\rho} = \xi_0 \rho^{s-1}.
\end{eqnarray}
Equations \eqref{xi} and \eqref{tau} are the two main hypotheses assumed in order to find a solution for the transport equation of the viscous pressure \eqref{transport IS}.

Combining the Friedmann equations \eqref{friedman 1} \eqref{friedman 2}, with the pressure in this context $p_{eff}=p+\Pi = w\rho +\Pi$, it is easy to obtain
\begin{equation}\label{Pi H}
    \Pi = -2\dot{H} -3(1+w)H^2.
\end{equation}
With the equation \eqref{Pi H}, a differential equation for the Hubble parameter can be constructed substituting \eqref{Temp IS}, \eqref{xi} and \eqref{tau} into the transport equation \eqref{transport IS}, then differentiate \eqref{Pi H} and equal both expressions considering the energy density conservation equation \eqref{conservation eq IS}.
The resulting equation for the Hubble parameter after these considerations is \cite{Cataldo:2005qh}: 
\begin{eqnarray}\label{H IS}
\ddot{H} &+& 3H \dot{H} +\frac{3^{1-s}}{\xi_0}\dot{H}H^{2-2s} -\left(\frac{1+2w}{1+w}\right)\frac{\dot{H}^2}{H} \nonumber \\
&+&\frac94(w-1)H^3 +\frac{3^{2-s}(1+w)}{2\xi_0}H^{4-2s} = 0.
\end{eqnarray}
There is not a general analytical solution to the last equation, it may be solved numerically for specific cases of the parameters of the model. 
In \cite{Cruz:2016rqi} the authors solved this equation for several values of $s$ numerically. 
The authors found that for $s \neq 1/2$ (or $s \leq -1/2$) there is no phantom solution.

However, for the particular case of $s = 1/2$, i.e. $\xi=\xi_0 \sqrt{\rho}$, the equation \eqref{H IS} has a particular analytical solution. 
Under this assumption the equation reduces to
\begin{eqnarray}\label{H IS s 1/2}
\ddot{H} &+& \left(3 +\frac{\sqrt{3}}{\xi_0}\right)H \dot{H} -\left(\frac{1+2w}{1+w}\right)\frac{\dot{H}^2}{H} \nonumber \\
&+&\frac94\left(w-1 +\frac{2(1+w)}{\sqrt{3}\xi_0}\right)H^{3} = 0.
\end{eqnarray}
Rewriting the last equation we have
\begin{eqnarray}\label{H IS s 1/2}
&\xi_0&(w+1)\ddot{H} + (w+1)\left(3\xi_0 +\sqrt{3}\right)H \dot{H} -\xi_0(1+2w)\frac{\dot{H}^2}{H} \nonumber \\
&+& \frac34(w+1)\left(3\xi_0(w-1)+2\sqrt{3}(1+w)\right)H^{3} = 0.
\end{eqnarray}
Notice that clearly if the bulk viscosity parameter is zero $\xi_0=0$, then (from the transport equation \eqref{transport IS}) the bulk viscosity pressure is null $\Pi=0$,  and we recover the standard Hubble equation
\begin{eqnarray}
\dot{H} + \frac32(1+w)H^2 = 0.
\end{eqnarray}
For the cosmological constant case, $w=-1$, equation \eqref{H IS s 1/2} boils down into the well-known de Sitter case $\dot{H} = 0$.
Also notice that under all the hypothesis considered to derive the evolution equation \eqref{H IS s 1/2}, for the standard case $w=-1$, the evolution equation gives a solution in which the bulk viscosity does not appear regardless of how $\xi$, $\tau$ and $T$ are chosen. 

Henceforth we will assume $\xi_0 \neq 0$ and $w+1\neq0$ to to exclude the standard case. 
For mathematical convenience, we define 
\begin{eqnarray}
a &=& \frac94\left(w-1 +\frac{2(1+w)}{\sqrt{3}\xi_0}\right), \\
b &=& 3 +\frac{\sqrt{3}}{\xi_0},
\end{eqnarray}
then the equation \eqref{H IS s 1/2} transforms into
\begin{eqnarray}\label{H IS s 1/2 b}
\ddot{H} + bH\dot{H} -\left(\frac{1+2w}{1+w}\right)\frac{\dot{H}^2}{H} + aH^{3} = 0.
\end{eqnarray}
To sketch a solution, the following Ansatz is proposed for the Hubble function in the late future universe \cite{Cruz:2016rqi, Cataldo:2005qh, Cruz:2017bcv, Cruz:2019wbl}:
\begin{eqnarray}\label{H ansatz}
H(t) = A(t_s - t)^{-1},
\end{eqnarray}
where $t_s$ is a finite time in the future at which the Big Rip occurs and $A$ is a positive constant in order to describe an expanding universe. 
For $s \neq 1/2$ it can be seen by direct inspection that a more general Ansatz with the form, $H(t) = A(t_s - t)^p$ with $p < 0$, does not reduce the differential equation to an algebraic polynomial in $A$ \cite{Cruz:2016rqi}.

Substituting the Ansatz \eqref{H ansatz} into equation \eqref{H IS s 1/2 b} it is obtained a quadratic equation for the constant $A$
\begin{eqnarray}
    aA^2 +bA +\frac{1}{w+1} =0,
\end{eqnarray}
whose solution is 
\begin{eqnarray}
    A_{\pm} = \frac12\left(-\frac{b}{a} \pm \sqrt{\left(\frac{b}{a}\right)^2 -\frac{4}{a(w+1)}}\right).
\end{eqnarray}
Integrating \eqref{H ansatz} we obtain the scale factor as a function of time, \cite{Cruz:2016rqi, Cruz:2017bcv}
\begin{equation}\label{a IS 1/2}
    a(t) = \left(\frac{t_s-t}{t_s-t_0}\right)^{-A},
\end{equation}
where $t_0$ is the present time.
With this it can be easily calculated the number of particles
\begin{equation}\label{n IS 1/2}
   n(t) = n_0 \left(\frac{t_s-t}{t_s-t_0}\right)^{3A}.
\end{equation}
It is clear that at the time $t = t_s$ the size of the universe becomes infinite and the number density of particles goes to zero. 

Finally, with the Ansatz \eqref{H ansatz} we calculate the bulk viscous pressure: 
\begin{equation}\label{Pi IS 1/2}
    \Pi(t) = -A(2 + 3(1+w)A)(t_s -t)^{-2}.
\end{equation}
Notice that the bulk viscous pressure is divergent at the Big Rip time.

\subsection{Temperature}

To calculate the expression of the temperature as a function of the scale factor we use the temperature as a function of the energy density equation \eqref{Temp IS}, and the equation of the Hubble parameter \eqref{H ansatz} and the scale factor \eqref{a IS 1/2}: 
\begin{eqnarray}\label{temp IS 1/2}
    T(t) &=& \tilde{T} (3A^2)^{w/(w+1)} \, (t_s-t)^{-2w/(w+1)} \label{Temp t IS 1/2}\\
    &=& \tilde{T} \left(3A^2 (t_s -t_0)^{-2}\right)^{w/(w+1)} \, a^{2w/A(w+1)} \nonumber \\
    &=& T_0  \, a^{2w/A(w+1)}, \label{Temp a IS 1/2}
\end{eqnarray}
with $T_0$ the temperature today.
It is a power-law in terms of the scale factor.
For the quintessence case, $-1 < w < -1/3$, the power is always negative, giving a temperature that decrease as the universe expands; whereas for phantom dark energy the power is positive giving an increasing temperature.
Once again, the characteristic behavior where the temperature increases as a function of the scale factor, becomes present.

\subsection{Entropy}

The entropy change can be evaluated from \eqref{entropy viscous}, $\dot{S} = -3H \Pi /(nT)$; substituting the Hubble Ansatz \eqref{H ansatz}, the viscous pressure \eqref{Pi IS 1/2}, the number of particles \eqref{n IS 1/2} and temperature as a function of time \eqref{Temp t IS 1/2}, we have: 
\begin{eqnarray}\label{dS/dt}
    \frac{dS}{dt} &=& K (t_s -t)^{\eta}, \\
    K &=& \frac{(3A^2)^{\frac{1}{w+1}} [2+3A(1+w)]}{n_0 T_0} (t_s -t_0)^{3A},\\
    \eta &=& \frac{2w}{w+1} -3(1+A).
\end{eqnarray}
To calculate the entropy we integrate \eqref{dS/dt} 
\begin{eqnarray}
    S(t) = - \left(\frac{K}{\eta+1}\right) (t_s -t)^{\eta+1} +\tilde{S},
\end{eqnarray}
where $\tilde{S}$ is an integration constant. We have a positive entropy $S > 0$, if $K>0$ and $\eta < -1$, that is, when $\frac{2w}{w+1} -3(1+A) < -1$, which leads to the a constraint for the constant $A$.
\begin{equation}
    A > \frac13 \left(1+\frac{2w}{w+1}\right) -1,
\end{equation}
and since $A > 0$ then, the last inequality is satisfied for values $0 < w < 1/2$. 
This is an interesting result, in the Israel-Stewart framework, for the particular case $\xi=\xi_0 \rho^{1/2}$; the natural conditions of positive entropy ($\eta+1<0$) and expanding universe ($A>0$) imply that the equation of state parameter must be positive, $w>0$.
This can be interpreted as the fluid is not a ``barotropic dark energy" ($w<-1/3$), but, it is important to take in mind that we have discussed the thermodynamic properties of a Big Rip solution $H=A(t_s-t)^{-1}$ in the framework of the Israel-Stewart theory and despite it gives $w>0$, the cosmic expansion is accelerating due to the global effect of the bulk viscosity in which $w_{eff} = p_{eff}/\rho< -1/3$.
The solution is for a cosmological scenario with barotropic fluid $p=w\rho$ and the constriction $0 < w < 1/2$, but, globally behaves like a phantom fluid, due to the bulk viscosity provides the sufficient negative pressure, allowing to cross the phantom divide line.

\section{Conclusions}\label{conclusions}

We have discussed in the present work a general treatment for dark energy thermodynamics in several scenarios for the late future time: by considering it as a perfect fluid with constant and variable equation of state parameter $w$, and as a dissipative fluid with bulk viscosity both in the Eckark and Israel-Stewart frameworks.
General equations have been derived for the dark energy temperature, energy density and entropy in a flat, homogeneous and isotropic universe.
From the results of the entropy evolution, some theoretical thermodynamic constraints are imposed for the $w$ parameter in order to satisfy a positive the entropy and grow rate. 

We first recall some main results of dark energy as a perfect fluid with constant $w$; previously presented in a clear way in the work  \cite{Duarte:2018gmt}.
In this context the temperature is directly proportional to the internal energy $T \propto U$, the energy density is a power-law of the temperature $\rho \propto T^{(1+w)/w}$, the entropy is constant in time (adiabatic) $\dot{S}=0$, and if the chemical potential is null $\mu=0$ then the phantom regime is forbidden ($w\geq -1$).
Then, we apply the same treatment for the dynamical dark energy case $w(a)$.
The reader should note two interesting points: one is that despite the dynamical character of $w$ there is no entropy production $S=constant$; and second, the chemical potential plays an important role, if it is null then necessarily $w(a) =w=const. \geq -1$, i.e, is not dynamical, if it is positive the phantom regime is avoided, and if it is negative the dark energy could be either quintessence-like ($w>-1$) or phantom-like ($w<-1$).

In the second part of this work, the dark energy is modeled as a dissipative fluid with bulk viscosity where the (effective) pressure is assumed as the sum of the barotropic and the bulk viscous pressures $p_{eff} = p +\Pi$.
This is the core of this work. 
Within the Eckart framework the viscous pressure is given by $\Pi = -3H\xi(t)$ and two cases were studied: $\xi(t) =\xi_0=const.$ and $\xi= (\beta/\sqrt{3})\rho^{1/2}$.
Both cases have the relation $T \propto \rho^{w/(1+w)}$.
In the first case, the entropy grows exponentially and puts the condition $w>-1$ to preserve the entropy and temperature positive, or, if $w>-1$ then $T<0$.
Although the possibility of negative temperature for phantom energy has been studied \cite{Saridakis:2009uu}, it seems a cumbersome hypothesis and therefore quintessence is favored.
In the second case, the entropy grows as a power-law, and quite an interesting fact in this case is that the phantom regime is allowed under not so strict conditions.
The next step is to use the causal Israel-Stewart theory.
We have solved the bulk viscous pressure transport equation assuming $\xi=\xi_0 \rho^{1/2}$ and $\tau = \xi/\rho$ which for the late future time universe gives rise to Big Rip cosmological solutions, $H=A(t_s-t)^{-1}$ \cite{Cruz:2016rqi, Cataldo:2005qh, Cruz:2017bcv}.
Under these particular assumptions the temperature evolves as $T = T_0\, a^{2w/A(w+1)}$, that is, in the quintessence regime the temperature decreases as the universe expands, whereas the temperature increases in the phantom regime. 
This is a characteristic of phantom and Big Rip solutions, here the energy density increases with the cosmic expansion as well the internal energy and consequently the temperature too.
Calculating the entropy we obtain a power-law in time and imposing the conditions $A>0$ and $S>0$, an important constraint $0< w <1/2$ is derived.
The solution implies a cosmological scenario of a barotropic fluid with $0 < w < 1/2$ which behaves like a dark fluid (even phantom) driven by the bulk viscosity.
So, in the Israel-Stewart formalism, some solutions behave as dark energy and even allow to cross the phantom divide line with an effective equation of state lesser than $-1$ and constant in time without evoking an exotic barotropic fluid with $w<-1$. 
An important extension to the study done is to obtain other cosmological solutions additional to $H=A(t_s-t)^{-1}$ and constrain the model parameters with observational data to track the most plausible model.

Despite its intrinsic nature is not well understood yet, reconstructions of the dark energy equation of state from observational data seem to suggest a time-dependent $w$ and even the crossing of the phantom divide line $w = -1$ from above to below is not only possible but could indeed be a condition for a successful description of observations. 
For this reason it is important to examine the phantom regime carefully.
It has important problems and if modeled as a perfect fluid, it seems unfeasible from the point of view of classical thermodynamics.
Hence, new perspectives beyond the perfect fluid approach in standard cosmology have to be considered in order to have a better understanding of the phantom regime, such as fluids with bulk viscosity that reduce the kinetic pressure of the cosmological fluid and provide a richer thermodynamic framework.
In fact, one could actually dive deeper and seek for the foundations of dark energy thermodynamics from statistical physics (and possibly quantum mechanics) to establish basic general principles for a dark energy theory, or if it is the case, give solid arguments in favor of alternatives like modified gravity avoiding the hypothesis of dark energy as a substance; all this important but long beyond the scope of this work.

Finally, dark energy thermodynamics and bulk viscous dark energy are topics that have been studied from several different approaches for years, in this work we summarize particular results previously obtained by other authors on both subjects to make a broader and unified study, extending its scope in order to search for insights about the nature of the dark energy from a general theory that is classical thermodynamics and more general fluids than the perfect fluid.

\begin{acknowledgments}
DT acknowledges the receipt of the grant from the Abdus Salam International Centre for Theoretical Physics, Trieste, Italy; and would like to thank Karen Caballero for the hospitality provided at the Facultad de Ciencias en F\'isica y Matem\'aticas, Universidad Aut\'onoma de Chiapas.
\end{acknowledgments}

\end{document}